\documentclass[aps,twocolumn,
superscriptaddress,
footinbib,
pra]{revtex4-2}

\usepackage{mathtools}
\usepackage{bbold}
\usepackage{amsmath,amssymb,bm}
\usepackage{mathrsfs}
\usepackage{MnSymbol}
\usepackage{graphicx}
\usepackage{epstopdf}
\usepackage{latexsym}
\usepackage[normalem]{ulem}
\usepackage[caption=false]{subfig}
\usepackage[usenames,dvipsnames]{color}
\usepackage[hidelinks]{hyperref}
\usepackage[nameinlink]{cleveref}
\usepackage{natbib}
\usepackage{xcolor, colortbl}
\usepackage{physics}
\usepackage{multirow}
\usepackage{stackengine}
\usepackage{dsfont}
\usepackage{relsize}
\usepackage{float}
\usepackage{framed}
\usepackage[utf8]{inputenc}
\usepackage[acronym]{glossaries}
\glsdisablehyper
\usepackage{booktabs}

\crefname{figure}{Fig.}{Fig.}
\crefname{section}{Sec.}{Sec.}
\crefname{equation}{Eq.}{Eq.}

\begin{document}

\newacronym{dfnn}{DFNN}{dense feedforward neural network}
\newacronym{gat}{GAT}{graph attention network}
\newacronym{lstm}{LSTM}{long short-term memory}
\newacronym{mcmc}{MCMC}{Markov chain Monte Carlo}
\newacronym{mle}{MLE}{maximum likelihood estimation}
\newacronym{povms}{POVMs}{Positive operator valued measures}
\newacronym{povm}{POVM}{positive operator valued measure}
\newacronym{povm-pd}{POVM-PD}{POVM probability distribution}
\newacronym{rnn}{RNN}{recurrent neural network}
\newacronym{tfim}{TFIM}{transverse field Ising model}
\newacronym{hce}{HCE}{half chain entropy}
\newacronym{mi}{MI}{mutual information}
\newacronym{ood}{OOD}{Out-Of-Distribution}
\newacronym{id}{ID}{In-Distribution}

\title{Sample-efficient estimation of entanglement entropy through supervised learning}

\date{\today}

\author{Maximilian Rieger}
\affiliation{Kirchhoff-Institut f\"{u}r Physik, Universit\"{a}t Heidelberg, Im Neuenheimer Feld 227, 69120 Heidelberg, Germany}
\author{Moritz Reh}
\email{moritz.reh@kip.uni-heidelberg.de}
\affiliation{Kirchhoff-Institut f\"{u}r Physik, Universit\"{a}t Heidelberg, Im Neuenheimer Feld 227, 69120 Heidelberg, Germany}
\author{Martin G\"{a}rttner}
\email{martin.gaerttner@uni-jena.de}
\affiliation{Kirchhoff-Institut f\"{u}r Physik, Universit\"{a}t Heidelberg, Im Neuenheimer Feld 227, 69120 Heidelberg, Germany}
\affiliation{Physikalisches Institut, Universit\"at Heidelberg, Im Neuenheimer Feld 226, 69120 Heidelberg, Germany}
\affiliation{Institut f\"ur Theoretische Physik, Ruprecht-Karls-Universit\"at Heidelberg, Philosophenweg 16, 69120 Heidelberg, Germany}
\affiliation{Institute of Condensed Matter Theory and Optics, Friedrich-Schiller-University Jena, Max-Wien-Platz 1, 07743 Jena, Germany}

\begin{abstract}
We explore a supervised machine learning approach to estimate the entanglement entropy of multi-qubit systems from few experimental samples. We put a particular focus on estimating both aleatoric and epistemic uncertainty of the network's estimate and benchmark against the best known conventional estimation algorithms. For states that are contained in the training distribution, we observe convergence in a regime of sample sizes in which the baseline method fails to give correct estimates, while extrapolation only seems possible for regions close to the training regime. As a further application of our method, highly relevant for quantum simulation experiments, we estimate the quantum mutual information for non-unitary evolution by training our model on different noise strengths.
\end{abstract}

\maketitle

\section{Introduction}
The ultimate feat in probing the quantum nature of quantum many-body systems consists in understanding their entanglement properties \cite{Horodecki2009}. Insights into phenomena such as the thermalization of closed quantum systems and many-body localization are fundamentally linked to entanglement \cite{Calabrese2005, BASKO2006, Abanin2019}, meaning the coherent delocalization of information among system constituents. The certification of entanglement is referred to as entanglement witnessing \cite{GUHNE2009, TERHAL2000}, and has been achieved in many systems of interest, including Bose-Einstein condensates \cite{Esteve2008}, photonic systems \cite{Mair2001}, atoms in optical lattices \cite{Dai2016} and many more. At the same time, the quantification of entanglement is disparately more challenging, as one na\"ively requires the (sub-)system's density matrix $\rho$ in order to compute the von-Neumann entropy $S=-\tr\left(\rho \ln \rho\right)$ or its Renyi-extensions, which is of interest due to its role as an indicator of quantum phase transitions \cite{Osborne2002} or its evolution in dynamical systems \cite{Calabrese2005}. However, as full tomography of $\rho$ becomes prohibitively expensive for larger systems due to the curse of dimensionality, it becomes increasingly challenging to obtain reliable estimates of entanglement entropies at reasonable experimental and computational costs \cite{Blume-Kohout2010, Lvovsky_2004}. One option to minimize these costs consists in restricting the functional form of the state to a certain type, decreasing the number of variational parameters at the expense of introducing a bias \cite{Cramer2010,Baumgratz2013,Lanyon2017,Gross2010,Schwemmer2014,Riofro2017,Toth2010,Moroder2012,Torlai2018,Cha2021,Carrasquilla2019, Zache2022, Joshi2023, Kokail2021, Anshu2021}. Notable asymptotically unbiased estimators that have been introduced to address these scaling issues include shadow tomography \cite{Huang} and distance-based approaches \cite{Elben}. While being more feasible than full scale tomography of the exponentially large density matrix $\rho$, the sample complexity of these methods is still too high, even for subsystems including only a small number of qubits, motivating the search for novel estimation schemes.

\begin{figure}
    \centering
    \includegraphics[width=0.45\textwidth]{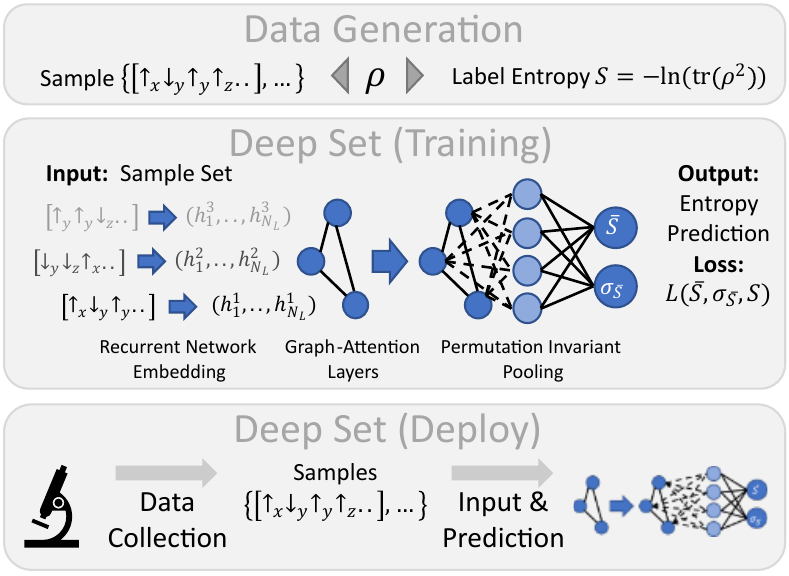}
    \caption{Visual description of the proposed procedure. We classically simulate a system of interest described by $\rho$, and store a quantity of interest, e.g. the entanglement entropy $S=-\ln \left(\tr\rho^2\right)$, which is difficult to estimate from samples. Simultaneously, we generate synthetic measurement data $\mathcal{S}$, which we use in a next step to learn a function that maps $\mathcal{S}$ to $S$. The function we propose for this is a deep neural network, that, in a first step, embeds all samples in a latent space, before mapping the set of latent space embeddings into another latent space, in a permutation invariant fashion. From there on, we use a feed-forward net to predict both the mean and the standard deviation of the estimate. Once trained, the model can be employed on unlabeled experimental data to give estimates for quantities with otherwise infeasible sample complexity, such as entanglement entropies.}
    \label{fig:visual_abstract}
\end{figure}

To overcome this challenge, we explore a supervised machine learning method that aims to estimate entropic quantities of quantum states, given a small set of samples. The approach is illustrated in \cref{fig:visual_abstract}. 
We are motivated by recent scientific breakthroughs based on supervised learning such as AlphaFold \cite{Jumper2021}, which demonstrate the capability of modern deep learning algorithms to find seemingly intractable maps with high complexity and no apparent structure.
Supervised applications in the realm of quantum physics have for example been explored in \cite{Moller2021, vanNieuwenburg2017, Carrasquilla2017}. A common goal is to build a controlled estimator, meaning a reliable treatment of uncertainties of the network output, which is why we lay a particular emphasis on their estimation.
In contrast to related work by Koutn\'y et al.\ \cite{Koutn2023}, our aim is to use as few samples as possible, being inspired by interacting spin systems rather than photonic experiments, thereby presenting a complementary approach to the problem. Additionally, various works have studied the problem of entanglement \textit{classification} employing machine learning tools \cite{Lu2018, Harney2020, Harney2021, Asif2023}, while we focus on entanglement \textit{quantification}.

We show that the proposed machine learning method is able to outperform the more general method proposed in \cite{Elben} when estimating the second order Rényi entanglement entropy, provided that we compare them on the domain the network was trained on (in-distribution). We also examine the network's ability to generalize to new domains and find that extrapolation is expectedly challenging. Finally, to motivate the use of the proposed method also in experimental scenarios, we study the performance in the presence of noise of varying type and strength and demonstrate correct estimation of the quantum mutual information.

\section{Physical System} While the proposed approach is applicable to any classically simulatable quantum system, we focus on the 1D transverse field Ising model as a benchmark system.

Its Hamiltonian is given by
\begin{align}
	\label[equation]{Hamiltonian}
	H = -J\sum_{i=1}^{N}\sigma_z^{(i)}\sigma_z^{(i+1)}
	+ h\sum_{i=1}^N\sigma_x^{(i)},
\end{align}
where $J \geq 0$ and employ periodic boundary conditions.
We will generate entangled states by quenching the paramagnetic ground state, $\ket{\psi_+} = \ket{+}^{\otimes N}$,
to the critical point $J=h=1$. We will consider both unitary time evolution,
governed by the Schrödinger equation
\begin{align}
	i\pdv{}{t}\ket{\psi} = H\ket{\psi}\qc \ket{\psi(t=0)} = \ket{\psi_+}
\end{align}
and non-unitary time evolution, determined by the Lindblad master equation
\begin{align}
	\label[equation]{Lindblad}
	\pdv{\rho}{t} = &-i \comm{H}{\rho} + \gamma_z\sum_{i=1}^N
	\left(\sigma_z^{(i)} \rho \sigma_z^{(i)} - \rho\right) \nonumber \\
	&+ \gamma_-\sum_{i=1}^N \left(\sigma_-^{(i)} \rho \sigma_+^{(i)}
	- \frac{1}{2}\acomm{\sigma_+^{(i)}\sigma_-^{(i)}}{\rho}\right), \nonumber \\
	\rho(t=&\:0) = \ket{\psi_+}\bra{\psi_+}.
\end{align}
The jump operators model single qubit noise, where $\sigma_z$ corresponds to
dephasing and $\sigma_- = \frac{1}{2}(\sigma_x - i\sigma_y)$ to decay. The
noise strength is determined by the parameters $\gamma_z, \gamma_- \geq 0$.
More details regarding the examined physical states will be given in the
according sections.

\section{Theoretical concepts}
We divide the system under scrutiny into the two subsystems $A$ and $B$, forming a bipartition with each subsystem representing half of the spin chain.
For a quantum state $\rho$, we define the reduced density matrix
\begin{align}
	\rho_{A} = \Tr_{B}(\rho).
\end{align}
For pure states $\rho$, the (Rényi)-entanglement entropy of order $n>1$ is 
defined as
\begin{align}
	S^{(n)}(\rho_A) = \frac{1}{1-n}\ln(\Tr(\rho_A^n)),
\end{align}
and we will focus on the case $n=2$ to allow for straight-forward comparison 
with the conventional method proposed in \cite{Elben}.

However, we note that the herein developed method does not depend on this choice and can be applied to any quantity of interest. In fact, we will also estimate the quantum mutual information
\begin{align}
	I(\rho) = S^{(1)}(\rho_A) + S^{(1)}(\rho_{A^{\complement}}) - S^{(1)}(\rho),
 \label[equation]{mutualInformation}
\end{align}
in the case of dissipative dynamics, generating mixed states, for which subsystem Rényi entropies no longer constitute a faithful measure of quantum correlations.
Here $S^{(1)}$ is the von-Neumann entropy, which can be seen by taking the limit
\begin{align}
	S^{(1)}(\rho) \coloneqq \lim_{n\searrow 1}S^{(n)}(\rho) =
	-\Tr(\rho\ln(\rho)).
\end{align}
In the remainder of this work we will abbreviate the
second order Rényi entropy of the half chain as \acrshort{hce}
\emph{(\acrlong{hce})} and the quantum mutual information as \acrshort{mi}.

Our aim is to estimate the aforementioned quantities based on a set of projective measurements.
Such projective measurements, taken in a single basis configuration, only partly describe the quantum 
state at hand. To estimate entanglement entropies, one also requires phase information,
which can be obtained by measuring in different basis configurations. If one is able to 
unambiguously infer the quantum state from the expectation values of a set of measurement operators, the set is said to be 
informationally complete.
\acrfull{povms} \cite{Nielsen} formalize this setup.
A \acrshort{povm} $\mathcal{M}$ is defined to be a
finite set of positive self-adjoint operators
\begin{align}
	\mathcal{M} = \{M_1, M_2, ...\}\qc M_i \geq 0\qc M_i^\dagger = M_i
\end{align}
that sum to unity, $\sum_i M_i = \mathbb{1}$.
Each $M_i$ represents one possible measurement outcome, which is observed with probability
\begin{align}
	\label[equation]{Born}
	p(M_i) = \Tr(\rho M_i).
\end{align}

Motivated by the simplicity of its experimental realization we here choose the Pauli-4-POVM, that is described by the measurement operators
\begin{align}
 \label[equation]{povm_operators}
	&M_1 = \frac{1}{3}\ket{+}\bra{+} \nonumber \qc
	M_2 = \frac{1}{3}\ket{L}\bra{L} \nonumber \qc
	M_3 = \frac{1}{3}\ket{0}\bra{0} \nonumber \\
	&M_4 = \mathbb{1} - M_1 - M_2 - M_3
\end{align}
where $\ket{+}, \ket{L}, \ket{0}$ are the $+1$ eigenstates of the
$\sigma_x, \sigma_y, \sigma_z$ operators. Operationally, this POVM can be implemented by measuring in all three Pauli bases and subsuming the outcomes with eigenvalues -1 in each basis under $M_4$. The Pauli-4-POVM can be easily generalized for $N$ qubits, by taking $N$-fold tensor products of the aforementioned measurement operators. A quantum state $\rho$ of a spin chain can then be associated to a probability distribution $P_\rho$ of generalized measurement outcomes, labeled by multi-indices $a \in \{1, 2, 3, 4\}^N$, given by
\begin{align}
 \label[equation]{povm_probability}
	P_\rho(a) = \Tr\left(\rho\left[\bigotimes_{i=1}^NM_{a_i}\right]\right).
\end{align}

\section{Data Generation}
Given a set of $N_M$ measurement outcomes sampled from this probability distribution, the network will be trained to estimate quantities such as the
\acrshort{hce} and the \acrshort{mi} of the quantum state $\rho$.
In order to generate the training dataset, it is necessary to 
classically simulate the system.
Obviously, this limits the system size regimes in which the proposed method can operate. Nevertheless, it is a meaningful extension over previous approaches as there is a gap between system sizes for which unbiased estimators give reliable results and system sizes which may be classically simulated. This statement holds irrespective of the Hamiltonian of the system and can even be extended to other quantum simulation platforms besides spin-1/2s.
The dataset comprises samples from $N_s$ states that differ in their physical parameters, for example the evolution time $t$, and their associated label such as the \acrshort{hce} and \acrshort{mi}. For each state we sample $N_B$ batches of $N_M$ measurement
outcomes from the \acrlong{povm-pd} using a Markov chain Monte Carlo algorithm. 
The total shape of the training dataset is thus given by 
$(N_S, N_B, N_M, N, 4)$ 
where the last dimension arises from an additional one-hot encoding of the four possible single-spin measurement outcomes introduced in \cref{povm_operators}.

\section{Network Architecture}
The network structure can be divided into three building blocks illustrated in \cref{fig:visual_abstract}. First, all POVM samples are embedded in a latent space using a \acrfull{lstm}.
The advantage of using a recurrent architecture is its ability to process inputs of arbitrary length which allows one to utilize the same architecture for systems of different size, although we do not explicitly exploit this feature in the present work.
The data will then be further transformed 
by a fully connected, permutation equivariant \acrfull{gat}. 
Permutation invariance is obtained by summation of all nodes of the \acrshort{gat}. A permutation invariant architecture restricts the search space to the relevant domain, eliminating redundant degrees of freedom from the parameter space facilitating an easier optimization \cite{geometricDeepl}.

In the last step the output of the \acrshort{gat} is evaluated by a 
\acrfull{dfnn} with two output neurons $(\bar{S}, \sigma_{\bar{S}})$ which we 
interpret as the mean $\bar{S}$ of the quantity of interest and its
statistical error $\sigma_{\bar{S}}$ of the estimation.
A more detailed description of the network structure is given in Appendix \ref{appendix:A}.

\section{Training \& Uncertainty Quantification}
We will train the network in a supervised fashion and generate training data using exact dynamics, as described above. If one is interested in systems of larger size, one could also generate training data using approximate methods, such as tensor networks \cite{Schollwoeck2011, Orus2014} or neural quantum states \cite{Carleo2017} by employing the replica trick \cite{Hastings2010}.

We define the loss as
\begin{align}
	L(\bar{S}, \sigma_{\bar{S}}; S; \theta) = 
	\frac{(S - \bar{S})^2}{\sigma_{\bar{S}}^2} + \ln(\sigma_{\bar{S}}^2),
\end{align}
where $S$ is the label, i.e.\ the true Rényi-entropy and $\theta$ denotes the vector of network parameters. The loss takes the form of the 
negative log-likelihood of a Gaussian distribution, from which the 
standard $L_2$ loss is recovered by setting $\sigma_{\bar{S}}$ to 1 \cite{aleatoricEpistemic}.
The uncertainty captured by $\sigma_{\bar{S}}$
is an estimate of the aleatoric uncertainty \cite{Huellermeier2021} and stems from the limited information on the probability distribution that can be inferred from the finite set of samples contained in the input. 
The aleatoric uncertainty is thus an inherent 
property of the dataset and cannot be reduced by means of a larger training dataset or similar.

By contrast, the reducible part of the total uncertainty which is 
determined by the choice of model, training scheme, etc. is usually called 
epistemic uncertainty. We attempt to estimate the epistemic
error by individually training an ensemble of $M$ networks with the same 
structure,
yet different initial parameters,
and averaging their outputs, after training is finished. This approach is
especially effective if the network is validated on domains that it did not
encounter during training, called \acrfull{ood}.
Referring to the outputs of model $m$ as $(\bar{S}_m, \sigma_{\bar{S}_m})$, the total uncertainty of the ensemble is given by \cite{aleatoricEpistemic}
\begin{equation}
    \sigma_{\bar{S}}^2 = \frac{1}{M}\left(\sum_{m=1}^M \sigma_{\bar{S}_m}^2 + \bar{S}_m^2 \right) - \left(\frac{1}{M} \sum_{m=1}^M \bar{S}_m \right)^2.
\end{equation}

The total computational cost of the approach is divided between training data generation and training of the network. In the instance of Fig.~\ref{lindblad1}, the generation of the data set took roughly 6 hours on a single NVIDIA A100 GPU, whereas the training of the network took a little more than 3 hours, which are representative values for all examples reported in this work.

\section{Results}
\subsection{Unitary Evolution}
The first application we consider is to predict the 
\acrshort{hce} of states obtained via unitary time evolution of the 
initial state $\ket{\psi_+}$ with parameters $J=h=1$. We set the system size to $N=10$ and train the network 
on $N_S = 100$ quantum states that were
taken at different points of the time interval $ht \in [0, 5]$. The batch size and sample size were chosen to 
be $N_B = 50$ and $N_M = 1000$. 
We validate the network during training on an independent dataset 
with the same parameters.
\begin{figure}[t]
	\includegraphics[width=0.5\textwidth]{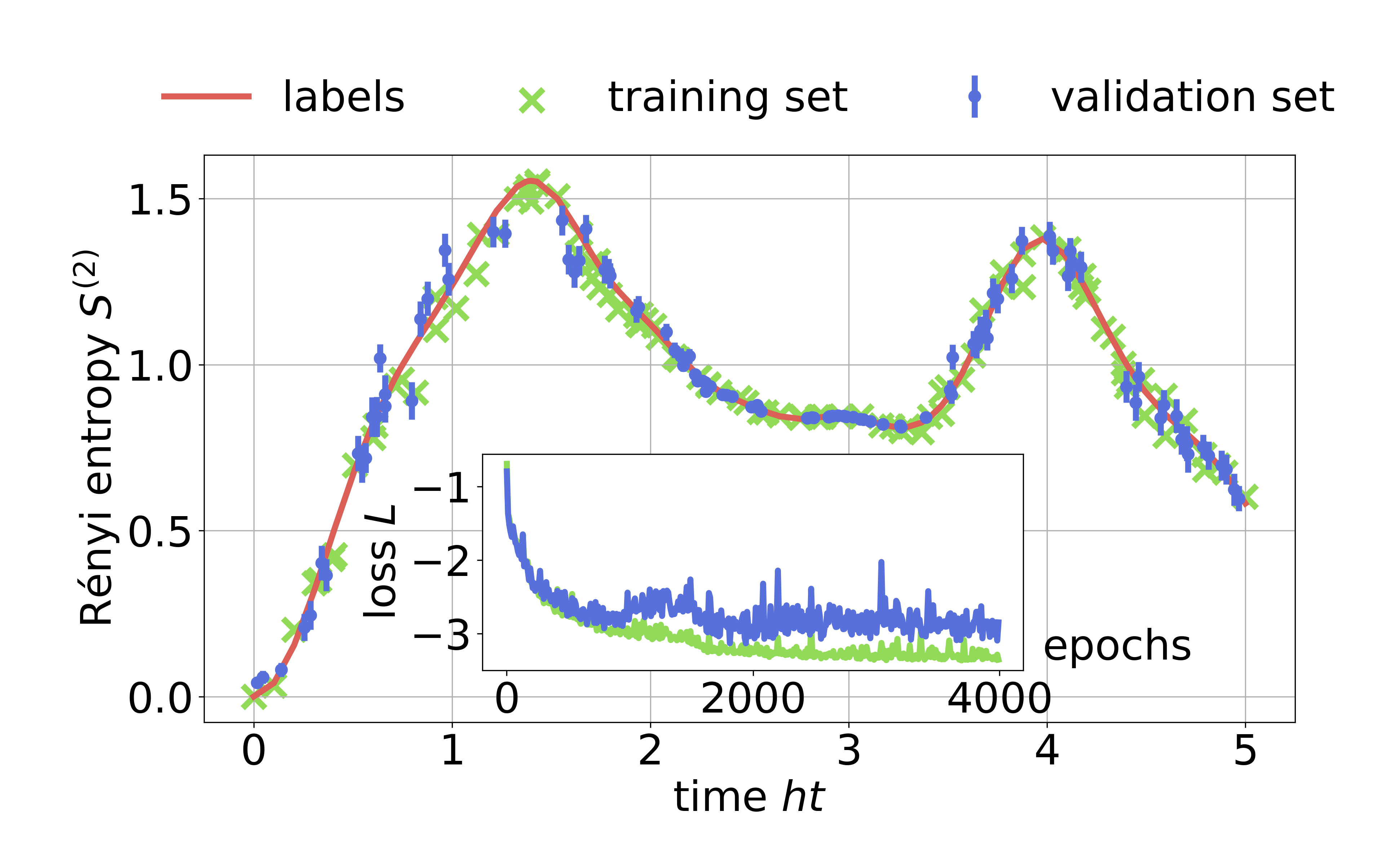}
	\caption{Training the network for $N = 10$.
	\emph{Main}: Networks performance
	on the training (\emph{green}, crosses) and validation (\emph{blue}, circles) dataset, 
	compared to training labels (\emph{red}, line).
	After 4000 epochs of training, the parameters with minimal loss on the
	validation dataset are selected.
	\emph{Inset}: Loss on training (\emph{green}, lower) and validation 
	(\emph{blue}, upper) dataset.}
	\label[figure]{unitary1}
\end{figure}
In \cref{unitary1} the network's predictions and the evolution of the loss during training are shown. It is apparent that the
network is capable of correctly estimating the \acrshort{hce} for a rather small dataset size.
The \acrshort{hce} is shown as a function of time for convenience,
but we want to emphasize that the network neither has information on the 
particular quantum state, nor the time at which the samples have been 
extracted from the unitary evolution. Predictions are solely based on 
\acrshort{povm} measurement results. 

\subsection{Extrapolation \& Baseline Comparison}
\begin{figure}[t]
	\includegraphics[width=0.45\textwidth]{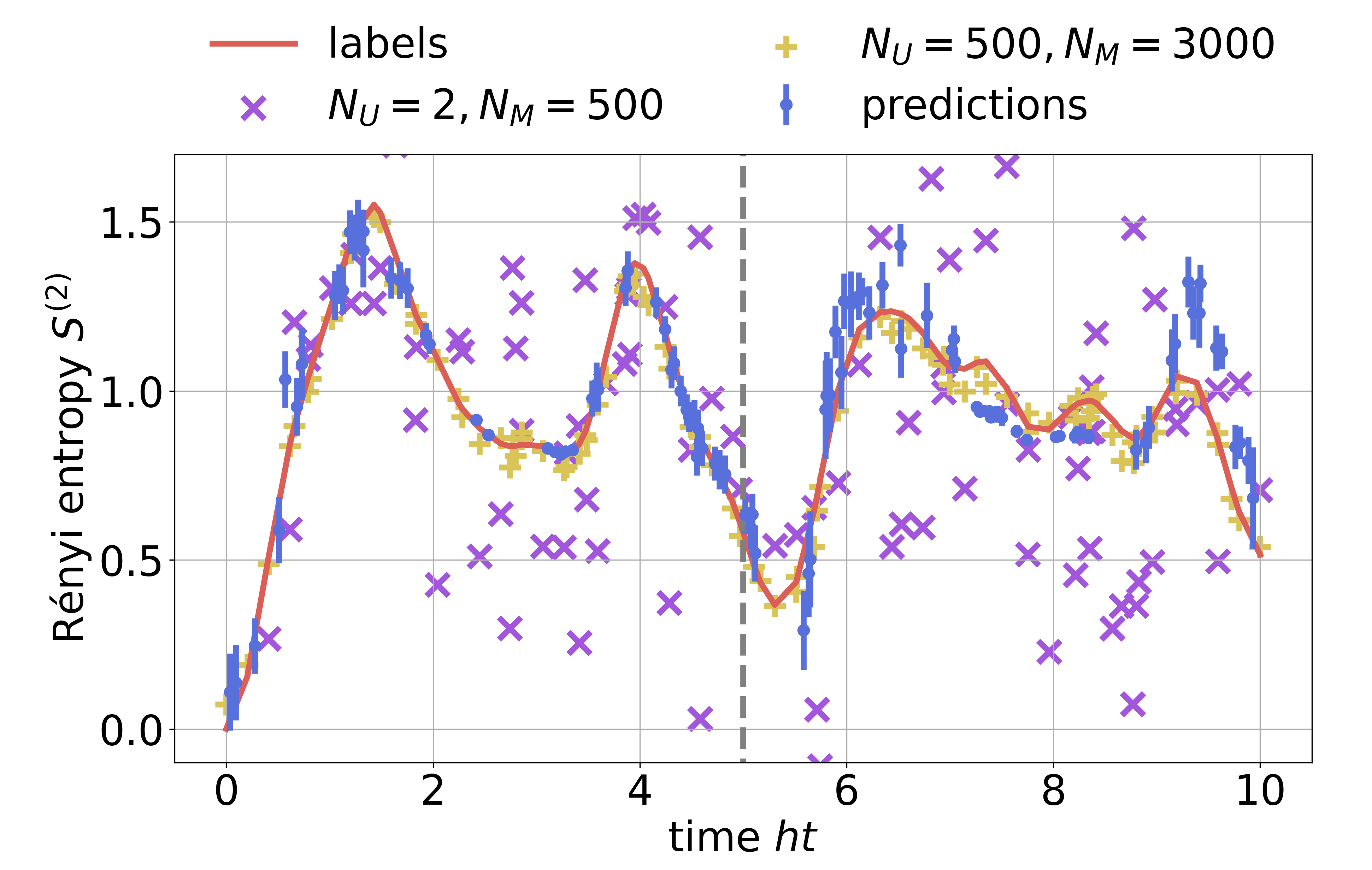}
	\caption{Performance of the network (\emph{blue}, circles) trained only on the 
	left hand side of the time interval (dashed \emph{grey} line).
	\emph{Violet} crosses: Baseline method for $N_U = 2$, $N_M = 500$.
	\emph{Yellow} pluses: Baseline method for $N_U = 300$, $N_M = 5000$.}
	\label[figure]{unitary2}
\end{figure}

We have demonstrated that the network succeeds at an interpolation task, when 
validated on \acrfull{id} data. We now aim to investigate the network's ability to estimate entanglement entropy on \acrshort{ood} data.
To this end we have trained an ensemble of 8 networks using the same hyperparameters 
and training dataset as in the previous case, but different initializations, and evaluated their performance on
quantum states on the larger time interval $ht \in [0, 10]$ as shown in \cref{unitary2}.
The predictions of the network seem sensible even beyond the training region up to $ht \approx 7$.
In the regime $ht \in [7, 8.5]$ however the error bars are severely
underestimated.

In order to obtain a better estimate of the performance enhancement that our method gives, we here compare to the method proposed in \cite{Elben}, which we use as a baseline.
It is based on
$N_M$ projective measurements, taken with respect to
$N_U$ randomly chosen orthonormal bases, such that a total
of $N_U \cdot N_M$ measurements have to be performed. From the measurement
statistics the subsystem purity $\tr(\rho_A^2)$ and thus the
\acrshort{hce} of the system may be inferred as shown in \cref{unitary2} for two different choices of hyperparameters.
In the first case the predictions are based on a total of 1000 measurements per
state, using the same number of samples as the network. In the second case the total
amount of measurements is $1.5\cdot 10^6$. One can see that the network
requires orders of magnitude fewer samples than the baseline to
properly predict the \acrshort{hce}, if it is evaluated on \acrshort{id} data. 
For \acrshort{ood} data it still outperforms the baseline, if evaluated for 
1000 measurements, but has less predictive power
compared to the baseline at $1.5\cdot 10^6$ samples, which is however a prohibitively large number of samples for current quantum simulation experiments. 

\begin{figure*}[t]
	\includegraphics[width=1\textwidth]{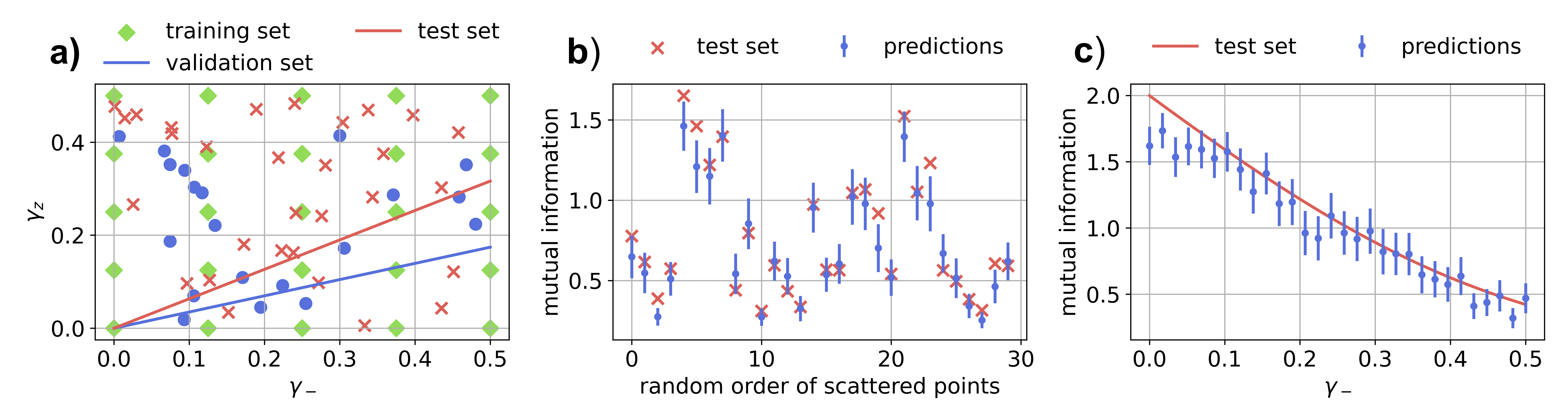}
	\caption{\textbf{a)}
	Training and validation datapoints in the $(\gamma_z, \gamma_-)$-plane.
	\emph{Green (diamonds):} Training set. 
	\emph{Blue (circles):} Validation set, consisting of 20 randomly distributed datapoints and 
	40 datapoints along a randomly chosen cross section (\emph{lower line}).
	\emph{Red (crosses):} Test set, consisting of 30 randomly distributed 
	datapoints and 30 validation datapoints along a randomly chosen 
	cross section (\emph{lower line}).
	\textbf{b)} Performance of network (\emph{blue circles}) 
	on randomly scattered datapoints (\emph{red crosses}). 
	\textbf{c)} Performance of the network (\emph{blue circles}) 
	on a cross section (\emph{red line}) of 30 datapoints.}
	\label[figure]{lindblad1}
\end{figure*}
\subsection{Dissipative dynamics}
If applied to noisy experimental data, robustness with respect to the experimental control and noise parameters is required. To demonstrate 
the usefulness of the proposed method in an actual experimental application, we now investigate the ability of the network to estimate the mutual information between the subsystems $A$ and $B$
in a dissipative setting.
We train the network on data obtained by modelling the Lindblad equation \eqref{Lindblad} with dephasing strength $\gamma_z$ and decay strength $\gamma_-$, considering a spin chain of length $N=8$ in the transverse field Ising model with initial state $\rho=\ket{\psi_+}\bra{\psi_+}$.
Instead of benchmarking on different evolution times, we instead limit ourselves to a fixed time $t=t_*=0.75$
and vary the noise strengths $\gamma_z$ and $\gamma_-$. 

We train the network on an equidistant $5 \times 5$ grid of noise strengths
$(\gamma_z, \gamma_-) \in [0, 0.5]^2$ with $N_B = 50$ batches of 
$N_M = 1000$ samples per
noise configuration. During training we validate the network on a dataset
sampled from quantum states with 20 randomly distributed noise configurations 
and 40 noise configurations which lie equidistantly spaced on a randomly
chosen cross section in the $(\gamma_z, \gamma_-)$-plane (\cref{lindblad1}a, \emph{blue}).
After training we selected the parameters which minimized 
the loss on the validation dataset and evaluated the performance of the network on a test set (\cref{lindblad1}b and \cref{lindblad1}c).

The network's predictions are overall close to the labels. We can therefore 
deduce that the network was indeed successful in learning to extimate the 
\acrshort{mi} for any noise strength. 
This underlines the applicability of the proposed method to experimental situations in which 
one often has a rough understanding of the underlying noise model, but no precise 
knowledge of the noise strengths.

\section{Discussion} We have demonstrated the possibility of learning maps from informationally complete POVM measurement data to quantum entropies using deep neural networks, allowing us to significantly reduce the sample complexity compared to the baseline procedure \cite{Elben}. While the estimator in \cite{Elben} is asymptotically unbiased, the approach that is followed here is not. Instead, we rely on training a model on labeled data and are therefore restricted to regimes that are classically simulatable. However, this still expands on the system size regimes that one may probe using the method proposed in \cite{Elben} since the regimes that allow for classical simulations comprise those for which one is able to generate sufficiently many samples for \cite{Elben} to give satisfactory results.

We tested the proposed method for both \acrshort{id} and \acrshort{ood} data, observing good performance for interpolation tasks. In extrapolation tasks, however, we find regimes in which all the networks contained in the ensemble predict similar entropies, resulting in uncertainties that are too small. However, since uncertainty quantification is a highly active research field, one can expect further advances in the estimation of epistemic uncertainties, which will be directly applicable to the method proposed here. In this context, an interesting future direction is to further explore the extrapolation capabilities, particularly in system size. Here, we wish to also point out that the proposed method is not limited to the estimation of entropies and mutual informations, but to any quantity and system for which both training data and labels can be generated synthetically. 

\vspace{0.5cm}
\textit{Code availability:} The full implementation of the network architecture is available through \href{https://github.com/MaximilianRie/EntropyEstimation}{GitHub}.

\acknowledgments
This work is supported by the Deutsche Forschungsgemeinschaft 
(DFG, German Research Foundation) under Germany’s Excellence Strategy 
EXC2181/1-390900948 (the Heidelberg STRUCTURES Excellence Cluster) and 
within the Collaborative Research Center SFB1225 (ISOQUANT). 
This work was partially financed by the Baden-Württemberg Stiftung gGmbH. 
The authors acknowledge support by the state of Baden-Württemberg through bwHPC
and the German Research Foundation (DFG) through Grant No INST 40/575-1 FUGG 
(JUSTUS 2 cluster). The authors gratefully acknowledge the Gauss Centre for 
Supercomputing e.V. (www.gauss-centre.eu) for funding this project by providing 
computing time through the John von Neumann Institute for Computing (NIC) on 
the GCS Supercomputer JUWELS 
\cite{JUWELS} at Jülich Supercomputing Centre (JSC).

\iftrue
\appendix

\begin{table}[t]
\begin{center}
\begin{tabular}{ ll }
\toprule
\textbf{Hyperparameter}\hspace{1cm} & \textbf{Value}\\\hline
RNN layers & 3 \\
RNN features ($F$) & 20, 20, 20\\
GAT layers & 2\\
GAT features $(F')$ & 10, 10\\
DFNN layers & 2\\
DFNN features & 4, 2\\
learning rate & 0.0005\\
\bottomrule
\end{tabular}
\end{center}
\caption{Network hyperparameters.}
\label{HypParams}
\end{table}

\section{Neural Network architecture}
\label{appendix:A}
The input of the neural network contains $N_M$ \acrshort{povm} 
samples, with shape $(N, 4)$ as they consist of $N$ single qubit 
measurement results with 4 possible outcomes (one-hot encoding, also see \cref{povm_probability}).
As explained in the architecture section of the main text, the network architecture is divided into three parts. At the first instance each sample is 
transformed by a layer of \acrshort{lstm} cells that iterates over the single qubit outcome, thereby mapping each sample into a latent space of dimension $F$. The embedded sample set therefore has dimension $(N_M, F)$. 
The embedded sample set is then treated by the \acrshort{gat} as a fully connected graph. During the development stages of the project we observed that a connected graph resulted in better accuracies compared to an unconnected graph, albeit incurring a much higher computational cost.
In each layer of the \acrshort{gat} the nodes are updated 
using self-attention \cite{GAT}. We choose to use two layers and a feature vector of dimension $F'$, which corresponds
to a transformation $(N_M, F) \mapsto (N_M, F')$ of the input. 
In order to obtain a 
permutation invariant quantity from the equivariant output of the 
\acrshort{gat} we sum over all $N_M$ nodes 
and receive a single invariant feature vector of dimension $F'$, that encapsulates all information of the sample set.
Finally, this feature vector is fed into a \acrshort{dfnn} with two output neurons which correspond to 
the prediction of the \acrshort{hce} and its aleatoric error.
As an optimizer for the parameters we have chosen ADAM \cite{Adam}.
The choice of network hyperparameters can be read from \cref{HypParams}.

\bibliography{refs}

\end{document}